\documentclass[preprint,aps,superscriptaddress,nofootinbib]{revtex4-2}

\usepackage{graphicx}

\usepackage{dcolumn}   
\usepackage{bm}       
\usepackage{amssymb}
\usepackage{amsmath}
\usepackage{blindtext}
\usepackage[T1]{fontenc}
\usepackage{amsmath}
\usepackage{mathtools}
\usepackage{amsfonts}
\usepackage{amssymb}
\usepackage{graphicx}

\newcommand{\del}{\partial}
\begin{document}
	
	\title{A Topological Origin of Black Hole Mass}
		\author{Sandipan Sengupta}
	\email{sandipan@phy.iitkgp.ac.in}
	\affiliation{Department of Physics, Indian Institute of Technology Kharagpur, Kharagpur-721302, India}
\begin{abstract}
We show that the notion of a black hole mass could be superceded by that of a topological charge in a spacetime without matter and curvature singularity. This feature emerges through a set new spacetime solutions of first order gravity in vacuum, named as bubble spacetimes, constructed here by weaving together the degenerate and nondegenerate metric phases. For a static bubble, the boundary surface connecting the two phases is characterized by a universal topological number. Notably, this surface coincides with the photon sphere of the conventional black hole, irrespective of the presence of the cosmological constant. In contrast, a phase boundary located at the event horizon is shown to be topologically trivial. Thus, along with the black hole mass, the photon sphere also acquires a topological interpretation.

\end{abstract}	

\maketitle 
\newpage

\section{Introduction}
The gravitational field outside a spherically symmetric mass distribution is given by the Schwarzschild solution \cite{schw} of General relativity. This geometry is parametrized by an integration constant, which reflects the total interior mass in the form of either an extended distribution or a point mass located at the spacelike singularity. In an idealized dynamical description of the formation of a black hole, the Schwarzschild metric defines the exterior of a collapsing FLRW spacetime with matter \cite{datt,os}. Within both the static and time-dependent scenarios above, the `Schwarzschild mass' necessarily originates due to genuine matter fields sourcing the curvature. 
Although the Schwarzschild spacetime is the simplest prototype of a black hole, this is a generic feature across such singular solutions of Einstein
gravity.

Possible departures within vacuum gravity from the conventional Schwarzschild black hole solution had been explored though \cite{bengtsson,varadarajan,kaul2017,sengupta2017}, within formulations where degenerate tetrad spacetimes arise as natural solutions to the field equations \cite{einstein,tseytlin,horowitz,jacobson,kaul2016,sengupta2016}. Bengtsson \cite{bengtsson} first demonstrated that an empty interior consisting of a degenerate set of densitized (spatial) triad fields could be joined to the Schwarzschild exterior solution such that the full spacetime is a solution to the Sen-Ashtekar constraints. Much later, based on a general method to construct degenerate spacetime solutions to the equations of motion of the Hilbert-Palatini Lagrangian density \cite{kaul2016,sengupta2016}, a different continuation of Schwarzschild exterior based on a set of degenerate tetrad (corresponding to a vanishing lapse function) had been conceived \cite{kaul2017}. These solutions happen to exhibit nonvanishing torsion in general. In both cases above, the event horizon plays a crucial role, which necessarily defines the location of the boundary hypersurface between the degenerate and nondegenerate phases. 

Motivated by some of these earlier studies, here we critically examine the question of the origin of the conventional black hole mass within a spacetime where the metric exhibits degeneracy at some region. In particular, we wonder if the source of this mass could be purely topological in the absence of matter and (geometric) torsion. While being quite contrary to the standard Einsteinian view of a black hole mass, such an investigation nevertheless reflects an intriguing perspective, and could open up a rich array of prospects.

To this end, first we present a general method for consistently gluing any spherically symmetric black hole exterior to a degenerate metric (in vacuum) across a dynamical phase boundary. This results in a continuous bubble geometry which could in general be time-dependent.
For the static solutions, the bubble mass, or equivalently, the apparent `black hole mass' in such spacetimes reflect a purely topological origin. To contrast with some of the earlier degenerate extensions \cite{bengtsson,kaul2017}, these geomeries have no torsion, the associated fields are real everywhere and the phase boundary does not coincide with the event horizon of the conventional black hole. 

Here we consider all possible classes of spherically symmetric black hole spacetimes in vacuum, namely, the Schwarzschild, Schwarzschild-dS and Schwarzschild-AdS geometries. In each case,  we find that the photon sphere appears as the unique boundary separating the degenerate and nondegenerate metric phases in the static solutions and corresponds to a universal topological charge. Thus, the bubble spacetimes provide a novel topological interpretation to the photon sphere as well to the Schwarzschild mass. The full spacetime in each case is constructed as a classical solution to the field equations of first-order gravity. This becomes possible due to the well-known fact that this formulation of gravity does not rely on the existence of an inverse metric (tetrad), thus being able to admit solutions with vanishing as well as non-vanishing metric (tetrad) determinant ($g$). We also show that that these geometries are associated with finite curvature scalars everywhere.

In the next section, we present the general details of our construction of the bubble spacetimes in vacuum. This is followed by an investigation of the underlying topological structure of the degenerate phase. Then we study the structure of the associated curvature scalars. The concluding section contains a few relevant remarks along with a discussion of future prospects.

\section{Topological bubble solutions}
We may recall the first-order gravity action with the cosmological constant:
\begin{eqnarray}\label{L}
{\cal L}(e,\omega)=\epsilon^{\mu\nu\alpha\beta}\epsilon_{IJKL} e_\mu^I e_\nu^J \Big[R_{\alpha\beta}^{~~KL}(\omega)~-~\frac{\Lambda}{3}e_\alpha^K e_\beta^L\Big]
\end{eqnarray}
The associated field equations are obtained by varying the above with respect to the connection $\omega_\mu^{~IJ}$ and the tetrad $e_\mu^I$ repectively:
\begin{eqnarray}\label{EOM1}
\epsilon^{\mu\nu\alpha\beta}\epsilon_{IJKL} e_\nu^K D_\alpha(\omega) e_\beta^L=0,\nonumber\\
\epsilon^{\mu\nu\alpha\beta}\epsilon_{IJKL} e_\nu^J \Big[R_{\alpha\beta}^{~~KL}(\omega)-\frac{2\Lambda}{3}e_\alpha^K e_\beta^L\Big]=0
\end{eqnarray}
This system becomes equivalent to Einstein's equations of General relativity only when the tetrad is invertible.

In this work, we shall restrict ourselves to the spherically symmetric black hole spacetime solutions of eq.(\ref{EOM1}). In general, these are solutions corresponding to the invertible phase ($g\neq 0$) of the system of equations (\ref{EOM1}) and are of the following form:
\begin{eqnarray}\label{bh}
	ds^2&=& -f(r) dt^2 ~+~ g(r) dr^2~+~ \chi^2(r)(d\theta^2+ \sin ^2\theta d\phi^2)\nonumber\\
	&&~
\end{eqnarray}
where $\chi'(r)\neq 0$.
We consider the following coordinate transformation $(t,r)\rightarrow(T,R)$: 
 \begin{eqnarray}\label{ct}
 t=t(T),~r=r(T,R),
\end{eqnarray}
 satisfying the condition $\frac{\del{r}}{\del R}|_{R_0}=0$ for some finite $R_0$. 
 At $R=R_0$, the transformed metric reduces to:
 \begin{eqnarray}\label{g}
	ds^2 &\doteq&   -\Bigg[f\Big(\frac{dt}{dT}\Big)^2-g\Big(\frac{d\rho}{dT}\Big)^2\Bigg] dT^2 
	+ \chi^2(\rho)(d\theta^2+ \sin ^2\theta d\phi^2)
\end{eqnarray}
where we define: $\rho(T)\equiv r(T,R_0)$. The symbol `$\doteq$' above implies equality only at $R=R_0$. 

In terms of the new coordinates, the field strength components read:
 	\begin{eqnarray}\label{R}
	R^{01}&=&-\frac{1}{2}\del_r\Big(f^{-\frac{1}{2}}g^{-\frac{1}{2}}\del_r f\Big)\dot{t}r'dT\wedge dR~\doteq~ 0,\nonumber\\
	R^{02}&=&-\frac{1}{2}f^{-\frac{1}{2}} g^{-1}\del_r f \del_r \chi~\dot{t} dT\wedge d\theta\nonumber\\
	&\doteq& 
	-\frac{1}{2}f^{-\frac{1}{2}} g^{-1}\del_\rho f \del_\rho \chi~\dot{t} dT\wedge d\theta
	\nonumber\\
	R^{03}&=&-\frac{1}{2}f^{-\frac{1}{2}} g^{-1}\del_r f \del_r \chi~ \dot{t} \sin\theta dT\wedge d\phi\nonumber\\
	&\doteq& 
	-\frac{1}{2}f^{-\frac{1}{2}} g^{-1}\del_\rho f \del_\rho \chi~\dot{t} \sin\theta dT\wedge d\phi
	\nonumber\\
	R^{12}&=&-\del_r\Big(g^{-\frac{1}{2}} \del_r \chi\Big) \Big(\dot{r} dT+r'dR\Big)\wedge d\theta \nonumber\\
	&\doteq& 
	-\del_\rho\Big(g^{-\frac{1}{2}} \del_\rho \chi\Big)\dot{\rho} dT\wedge d\theta
	\nonumber\\
	R^{23}&=&\Big(1-g^{-1}(\del_r \chi)^2\Big)\sin\theta d\theta\wedge d\phi\nonumber\\
	&\doteq & 
	\Big(1-g^{-1}(\del_\rho \chi)^2\Big)\sin\theta d\theta\wedge d\phi\nonumber\\
	R^{31}&=&-\del_r\Big(g^{-\frac{1}{2}} \del_r \chi\Big) d\phi \wedge  \Big(\dot{r} dT+r'dR\Big) \nonumber\\
	&\doteq& 
	-\del_\rho\Big(g^{-\frac{1}{2}} \del_\rho \chi\Big)\dot{\rho}\sin\theta d\phi \wedge dT
\end{eqnarray}
where the dot and prime denote derivatives with respect to $T$ and $R$, respectively.

Next, let us study if there exists a solution to eqs.(\ref{EOM1}) corresponding to the $g=0$ phase, which could be joined consistently to the exterior spacetime defined above across the surface $\rho(T)$. Since the metric, field-strength and torsion are the fundamental gauge-covariant fields in this context, such a construction must ensure their continuity across this evolving phase boundary. 
In general, degenerate space-time solutions of first order gravity are known to exhibit torsion whose origin is purely geometric. However, here we shall consider the simplest case of vanishing torsion everywhere in the full spacetime: $T_{\mu\nu}^{~~I} \equiv \frac{1}{2}D_{[\mu}(\omega)e_{\nu]}^{~~I} =0$.

To this end, let us consider the following non-invertible metric, to be glued to the metric (\ref{g}) across the phase boundary:
 \begin{equation}\label{g-0}
ds^2=-F^2(T) dT^2+0+\chi^2(\rho)(d\theta^2+\sin^2\theta d\phi^2)
\end{equation}
We assume that $F(T)$ and its derivatives at the phase boundary do not vanish.
Given the set of degenerate tetrad: $e^0=FdT,~e^1=0,~e^2=\chi d\theta,~e^3=\chi \sin\theta d \phi$, the most general solution for the spin-connection fields corresponding to vanishing torsion are obtained following the general results obtained in Ref.\cite{kaul2016}:
\begin{eqnarray*}
\omega^{ij}=\bar{\omega}^{ij},~\omega^{1i}=\lambda e^i,
\end{eqnarray*}
where $i\equiv(0,2,3)$,  $\bar{\omega}^{ij}$ denotes the torsionless spin-connection components. The constant $\lambda$, having the dimension of inverse length, needs to be introduced due to the different dimensions of the tetrad and connection fields. The resulting expressions for the connection are displayed below:
\begin{eqnarray}\label{connection}
\omega^{01}&=&-\lambda FdT,~
\omega^{02}=\frac{\dot{\chi}}{F}d\theta,~
\omega^{03}=\frac{\dot{\chi}}{F}\sin\theta d\phi, \nonumber\\
\omega^{12}&=& \lambda \chi d\theta,~
	\omega^{23}=  -\cos\theta d\phi,~
	\omega^{31}= -\lambda\chi\sin\theta d\phi \nonumber\\
	&&~
\end{eqnarray}
Using these, the field-strength components associated with the non-invertible phase are evaluated to be:
\begin{eqnarray}\label{R-0}
R^{1i}&=&0,~R^{02}=\Bigg[\del_T\Big(\frac{\dot{\chi}}{F}\Big)-\lambda^2\chi F\Bigg]dT\wedge d\theta,\nonumber\\
R^{03}&=&\Bigg[\del_T\Big(\frac{\dot{\chi}}{F}\Big)-\lambda^2\chi F\Bigg]\sin\theta dT\wedge d\phi,\nonumber\\
R^{23}&=&\Bigg[1-\lambda^2\chi^2+\Big(\frac{\dot{\chi}}{F}\Big)^2\Bigg]\sin\theta d\theta\wedge d\phi
\end{eqnarray}

We now proceed to construct an explicit solution to the first-order field equations for the full spacetime.

Since the connection fields (\ref{connection}) are torsionless by construction, the connection equations of motion given by the first set of (\ref{EOM1}) are trivially satisfied. Next, considering the ettrad equations of motion in (\ref{EOM1}), the $\mu=R,~I=1$ component leads to:
\begin{eqnarray}\label{mc}
&&\epsilon^{abc} \epsilon_{ijk}e_a^i \Big[R_{bc}^{~~jk}-\frac{2\Lambda}{3} e_b^j e_c^k\Big]~ =~0~=~ F\Bigg[\frac{2\chi}{F}
\del_T\Big(\frac{\dot{\chi}}{F}\Big) +\Big(\frac{\dot{\chi}}{F}\Big)^2-(3\lambda^2+\Lambda) \chi^2+1  \Bigg]
\end{eqnarray}
It is straightforward to check that the remaining components of the field equations are identically satisfied.

Let us now turn to the continuity conditions.
Inspection of the field-strength components in (\ref{R}) and (\ref{R-0}) implies:
\begin{eqnarray}\label{cont-R}
&& \dot{\chi}\doteq  0,\nonumber\\
&& \del_T\Big(\frac{\dot{\chi}}{F}\Big)-\lambda^2 \chi F \doteq  -\frac{1}{2} f^{-\frac{1}{2}}g^{-1}\del_\rho f \del_\rho\chi ~ \dot{t}\nonumber\\
&& \Big(\frac{\dot{\chi}}{F}\Big)^2-\lambda^2\chi^2 \doteq -g^{-1}(\del_\rho \chi)^2 
\end{eqnarray}
Further, continuity of the metrics (\ref{g}) and (\ref{g-0}) at $R=R_0$ leads to:
\begin{eqnarray}\label{cont-g}
F^2\doteq f\dot{t}^2-g\dot{\rho}^2
\end{eqnarray}
The constraints (\ref{cont-R}) and (\ref{cont-g}) are time-dependent. Hence, these could be satisfied either at a particular instant $T=T_0$, or, at all times. These two classes correspond to the non-static and static solutions, respectively. We consider the two cases separately below.

\subsection{Non-static bubble}
Given $F(T)\neq 0$, the equation of motion (\ref{mc}) in this case has the following solution:
\begin{equation}\label{F-soln}
	F = \frac{\dot{\chi}}{\sqrt{\Big(\lambda^2+\frac{\Lambda}{3}\Big) \chi^2-1+\frac{C}{\chi}}}
\end{equation}
where $C$ is the integration constant. The first equality in (\ref{cont-R}) implies that $\chi(T)$ and hence $r(T)$ has an extremum at $T=T_0$: $\rho(T_0)\equiv r_0$. The phase boundary in this case is the two-sphere at $T=T_0,~R=R_0$.
The other conditions in (\ref{cont-R})  lead to:
\begin{eqnarray}\label{lambda}
\lambda^2=\chi^{-2}(r_0)(\del_\rho \chi (r_0))^2\Big)g^{-1}(r_0)
\end{eqnarray} 
This fixes the integration constant in the solution (\ref{F-soln}) as: $C=\chi(r_0)\Big(1-g^{-1}(r_0)(\del_\rho \chi (r_0))^2\Big)$. The solution to eq.(\ref{cont-g}) is given by:
\begin{eqnarray}\label{F-0}
F(T_0)=f^{\frac{1}{2}}(r_0)\dot{t}(T_0)
\end{eqnarray}

Since $\lambda\neq 0,~F(T_0)\neq 0$, the above equation implies: $g^{-1}(r_0)\neq 0,~f(r_0)\neq 0$. Thus, the phase boundary in these solutions cannot occur at the location of the event horizon of a conventional black hole. Hence, these solutions are inequivalent to the earlier degenerate extensions of black hole spacetimes presented in the context of Sen-Ashtekar and Hilbert-Palatini formulations of gravity \cite{bengtsson,kaul2017}. 

\subsection{Static bubble}

Here we consider the other possibility that the boundary conditions (\ref{cont-R}) and (\ref{cont-g}) are satisfied at all times. This then implies:
\begin{eqnarray}\label{cont-static}
\rho(T)&=&const.\equiv r_0,\nonumber\\
 \lambda^2 &=& \frac{(\del_\rho\chi(r_0))^2}{\chi^2(r_0)}g^{-1}(r_0),\nonumber\\
F(T) &=& f(r_0)^{\frac{1}{2}}\dot{t}(T)\nonumber\\
&=&\frac{1}{2\lambda^2 r_0} f^{-\frac{1}{2}}(r_0)g^{-1}(r_0)\del_\rho f(r_0) \del_\rho \chi (r_0) \dot{t}(T)
\end{eqnarray}
As the first equation above shows, the radius of the two-sphere at the $g=0$ phase does not change with time. The last equality provides a solution for the location of the phase boundary, $r_0$ as:
\begin{eqnarray}
\frac{2f(r_0)\del_\rho \chi(r_0)}{r_0\del_\rho f(r_0)}~=~1
\end{eqnarray}

Finally, the solution to the equation of motion (\ref{mc}) reads:
\begin{eqnarray}\label{eom-soln}
(3\lambda^2+\Lambda) \chi^2(r_0)~=~1
\end{eqnarray}


\subsection{A nontrivial topological charge}

The spacetime at the $\det g_{\mu\nu}\neq 0$ in general is characterized by a non-vanishing mass parameter. On the other hand, the degenerate phase contains no pointlike or extended mass distribution. Torsion, since it vanishes, cannot source any geometric mass here as well.  Hence, it is natural to wonder what could be the origin of the bubble hole mass $M$. This mass is what would be interpreted as the black hole mass by an asymptotic observer. Here, we explore this question in detail.

Let us consider the $\mu=R,~I=1$ component of the Bianchi identity $\epsilon^{\mu\nu\alpha\beta}\epsilon_{IJKL}D_\nu R_{\alpha\beta}^{~KL}=0$:
\begin{eqnarray}
&&\epsilon^{abc}\epsilon_{ijk}D_a R_{bc}^{~jk}=0=\epsilon^{abc}\epsilon_{ijk}\Big[\hat{D}_a(\hat{\omega}) \hat{R}_{bc}^{~jk}+4M_a^j\hat{D}_b M_c^k\Big]\nonumber\\
&&\Rightarrow \hat{D}_a(\hat{\omega})\left[\epsilon^{abc}\epsilon_{ijk}\hat{M}_b^j M_c^k\right]=0
\end{eqnarray}
where $\hat{D}_a(\hat{\omega})$ denotes the covariant derivative with respect to the three-connection $\hat{\omega}_a^{ij}(\hat{e})$. In the above, we have used the following identities:
$R_{bc}^{~1k}=\hat{D}_{[b}M_{c]}^k$ and $\hat{D}_{[a}\hat{R}_{bc]}^{~ij}=0$ where $\hat{R}_{bc}^{~ij}$ is made up of purely the three-connection.

Projecting the above equation along the internal timelike unit vector $n^i\equiv(1,0,0)$ and after flipping the derivative, we obtain:
\begin{eqnarray}
\del_c\left[\epsilon^{abc}\epsilon_{ijk}M_a^i M_b^j n^k\right]~=~
\epsilon^{abc}\epsilon_{ijk}M_a^i M_b^j \hat{D}_c(\hat{\omega})n^k\nonumber\\
~
\end{eqnarray}
For static bubble solutions, the above reduces to:
\begin{eqnarray}
\del_c\left[\epsilon^{abc}\epsilon_{ijk}M_a^i M_b^j n^k\right]~=~0~=~\del_a j^a
\end{eqnarray}
where we define the conserved current as: $j^a\equiv \alpha(\lambda,\Lambda)\epsilon^{abc}\epsilon_{ijk}M_b^i M_c^j n^k$. Note that the factor $\alpha$ is introduced to account for the fact that the normalization of this current is ambiguous upto a constant. $\alpha$ could in general depend on the dimensionless constant $\frac{\Lambda}{\lambda^2}$.
Since this conserved current is independent of the metric and the equations of motion, and depends only on the boundary properties of the degenerate spacetime, it is manifestly topological.

~

Next, let us find the topological charge associated with this current:
\begin{eqnarray}\label{Q}
Q~=~\frac{1}{4\pi}\int_\Sigma d^2 x~j^T~=~\alpha\lambda^2 \chi^2(r_0)
\end{eqnarray}
$Q$ being topological, it could only be a pure number. Using eq.(\ref{eom-soln}), this is possible provided: $\alpha=C\Big(3+\frac{\Lambda}{\lambda^2}\Big)$. We may choose $C=1$, which fixes the normalization once and for all.


\section{The Schwarzschild bubble solution}
In this section, we consider first order gravity in vacuum with $\Lambda=0$. The unique solution for $g\neq 0$ is given by the Schwarzschild metric, for which we have:
	\begin{eqnarray*}
	f(r)~=~g^{-1}(r)~=~1-\frac{2M}{r},~\chi(r)=r 
\end{eqnarray*}
Clearly, in view of the coordinate transformation (\ref{ct}), the spacetime in $T,R$ coordinates is equivalent to the Schwarzschild geometry only locally (at $R>R_0$). 

Using eq.(\ref{R}), we obtain the field strength components as:
 	\begin{eqnarray}\label{R--}
	R^{01}&=&\frac{2M}{r^3}\dot{t}r'dT\wedge dR\doteq 0,\nonumber\\
	R^{02}&=&-\frac{M}{r^2}\left(1-\frac{2M}{r}\right)^{\frac{1}{2}}\dot{t} dT\wedge d\theta\nonumber\\
	&\doteq& 
	-\frac{M}{\rho^2}\left(1-\frac{2M}{\rho}\right)^{\frac{1}{2}}\dot{t} dT\wedge d\theta
	\nonumber\\
	R^{03}&=&-\frac{M}{r^2}\left(1-\frac{2M}{r}\right)^{\frac{1}{2}}\dot{t}\sin\theta dT\wedge d\phi\nonumber\\
	&\doteq &
	-\frac{M}{\rho^2}\left(1-\frac{2M}{\rho}\right)^{\frac{1}{2}}\dot{t}\sin\theta dT\wedge d\phi \nonumber\\
	R^{12}&=&-\frac{M}{r^2}\left(1-\frac{2M}{r}\right)^{-\frac{1}{2}}\Big(\dot{r} dT+r'dR\Big)\wedge d\theta \nonumber\\
	&\doteq& 
	-\frac{M}{\rho^2}\left(1-\frac{2M}{\rho}\right)^{-\frac{1}{2}}\dot{\rho} dT\wedge d\theta
	\nonumber\\
	R^{23}&=&\frac{2M}{r}\sin\theta d\theta\wedge d\phi\doteq 
	\frac{2M}{\rho}\sin\theta d\theta\wedge d\phi\nonumber\\
	R^{31}&=&\frac{M}{r^2}\left(1-\frac{2M}{r}\right)^{-\frac{1}{2}}\sin\theta\Big(\dot{r} dT+r'dR\Big)\wedge d\phi
\nonumber\\
&\doteq &
\frac{M}{\rho^2}\left(1-\frac{2M}{\rho}\right)^{-\frac{1}{2}}\sin\theta\dot{\rho} dT\wedge d\phi
\end{eqnarray}
Along with the metric, the field-strength above define the $g\neq 0$ phase of the full spacetime.

Let us consider the degenerate phase as discussed in Sec-II. For the non-static case, the general solution to the equation of motion (\ref{mc}) with $\Lambda=0$ reads:
\begin{equation}\label{F-soln--}
	F = \frac{\dot{\rho}}{\sqrt{\lambda^2 \rho^2-1+\frac{2M}{\rho}}}
\end{equation}
From the continuity requirements (\ref{cont-R}) and (\ref{cont-g}), we obtain:
\begin{eqnarray}\label{lambda--}
\lambda^2 &=&\frac{1}{r_0^2}\Big(1-\frac{2M}{r_0}\Big)\nonumber\\
F(T_0)&=&\Big(1-\frac{2M}{r_0}\Big)^{\frac{1}{2}}\dot{t}(T_0)
\end{eqnarray}
The above reflects the fact already emphasized earlier, that is, the phase boundary cannot be located at $r=2M$. 

Finally, inserting (\ref{F-0}) into the second equality in (\ref{cont-R}), we obtain:
\begin{eqnarray}\label{r-dot}
\ddot{\rho}(T_0)=\frac{(r_0-3M)F^2(T_0)}{r_0^2}
\end{eqnarray}
This constraint unravels the possible classes of geometries that could arise in this context. For $2M<r_0<3M$ and $r_0>3M$, the radius $\rho(T)$ exhibits a maximum and minimum at the phase boundary, respectively. The critical case $r_0=3M$ leads to the static solutions, which is what we are interested in here. These are discussed next.

The solutions to eqs. (\ref{cont-static})-(\ref{eom-soln}) corresponding to the time-independent case finally imply:
\begin{eqnarray}\label{lambda-sol}
\rho(T)~&=&~r_0~=~3M,\nonumber\\
\lambda~&=&~\frac{1}{r_0}\Big(1-\frac{2M}{r_0}\Big)^{\frac{1}{2}}=\frac{1}{3\sqrt{3}M},\nonumber\\
F(T)~&=&~\frac{1}{\sqrt{3}}\dot{t}
\end{eqnarray}
The resulting solution for the metric at the $g=0$ phase reads:
\begin{eqnarray}
ds^2~=~-\frac{1}{3}\dot{t}^2 dT^2+0+r_0^2(d\theta^2+\sin^2\theta d \phi^2)
\end{eqnarray}
Since $t(T)$ is arbitrary (with $\dot{t}(T)\neq 0$), we may choose: $t(T)=T$.

For this static Schwarzschild bubble, $r_0=3M$ reflects the unique radius at which a phase boundary could be located. Notably, this coincides with the photon sphere of a standard Schwarzschild black hole. This is precisely the critical case in (\ref{r-dot}).

\subsection*{Topological interpretation of Schwarzschild mass}

Based on the above solution, we now show that the static solution reflects a nontrivial topological charge. Using ($\ref{Q}$), this is evaluated to be: 
\begin{eqnarray}\label{q}
Q=3\lambda^2 r_0^2=1.
\end{eqnarray}
Further, using the second equation in (\ref{lambda-sol}), we
may write:
\begin{eqnarray}\label{m-top}
\frac{M}{r_0}=\frac{1}{2}\Big(1-\frac{Q}{3}\Big)
\end{eqnarray}
Clearly, the Schwarzschild `mass' (measured in units of $r_0$) corresponding to the bubble solution  has a topological origin.
The value $Q=1$ as obtained in (\ref{q}) earlier reflects the nontrivial topological charge of the Schwarzschild vacuum bubble. Equivalently, this encodes the topological number of the 
phase boundary which coincides with the photon sphere of the standard black hole.

From (\ref{m-top}), we also note that a phase boundary located at the event horizon $r_0=2M$ of a conventional black hole, as considered in some of the earlier works \cite{bengtsson,kaul2017}, is topologically trivial. This unravels an important contrast between the topological features of the photon sphere and the event horizon.

\subsection*{Regularity of curvature scalars}

The spacetime (\ref{g}) with $g\neq 0$ at $R>R_0$ is characterized by finite curvature scalars, since this is locally equivalent to the Schwarzschild exterior. At the non-invertible metric phase, the effective spacetime is three-dimensional. We now study the structure of the curvature scalars of the latter.
 
 Given the effective three-metric, we may define a set of invertible triad $\hat{e}_a^i\equiv e_a^i$ ($i\equiv[0,2,3],~a\equiv[T,\theta,\phi]$). Their inverse are denoted by $\hat{e}^a_i$ ($\hat{e}^a_i \hat{e}_{aj}=\eta_{ij},~\hat{e}^b_k \hat{e}_a^k=\delta_a^b$). The three dimensional internal metric is given by: $\eta_{ij}\equiv Diag[-1,1,1]$. From (\ref{g-0}), these triad fields could be read off as:
 \begin{eqnarray*}
 \hat{e}^0=F(T)dT,~\hat{e}^2=\rho(T) d\theta,~\hat{e}^3=\rho(T) \sin\theta d\phi.
 \end{eqnarray*}
With these, the set of torsionless connection fields $\hat{\omega}^{ij}(\hat{e})$ are evaluated to be:
 \begin{eqnarray*}\label{connection1}
\hat{\omega}^{02}=\frac{\dot{\rho}}{F}d\theta,~	\hat{\omega}^{03}=\frac{\dot{\rho}}{F}\sin\theta d\phi, ~\hat{\omega}^{23}=  -\cos\theta d\phi.
\end{eqnarray*}
The non-vanishing components of the resulting Riemann curvature tensor in the coordinate basis read:
\begin{eqnarray}
\hat{R}_{T\theta}^{~T\theta}&=&\lambda^2-\frac{M}{\rho^3}=\hat{R}_{T\phi}^{~T\phi},\nonumber\\
\hat{R}_{\theta\phi}^{~\theta\phi}&=&\lambda^2+\frac{2M}{\rho^3}
\end{eqnarray}
where $\lambda$ is given by (\ref{lambda}).
From the above, we conclude that the non-static solutions where the radius $\rho(T)$ exhibits a minimum at the phase boundary $T=T_0$ exhibit no curvature singularity in its full range. This is so because the Riemann components displayed above, whose products essentially define all possible curvature scalars here, are finite. 

Similarly, the static solutions with $\rho(T)=3M$ are also manifestly free of curvature singularity, $
\hat{R}_{\theta\phi}^{~\theta\phi}=\frac{1}{9M^2}$ being the only nontrivial component.

The remaining (non-static) class where $\rho(T_0)$ is a maximum of $\rho(T)$ is manifestly singular, and hence would be discarded.

\section{Schwarzschild-de Sitter (Anti de Sitter) bubble solution}

For a nonvanishing $\Lambda$, the unique solution to vacuum gravity is given by:
\begin{eqnarray}\label{sds}
f(r)=g^{-1}(r)=1-\frac{2M}{r}-\frac{\Lambda r^2}{3},~\chi(r)=r.
\end{eqnarray}
The cases $\Lambda>0$ and $\Lambda<0$ correspond to the Schwarzschild-dS and Schwarzschild-AdS metric, respectively.
Eq.(\ref{sds}) defines the geometry of the spacetime at $R>R_0$. Since the analysis is similar to the earlier section in outline, we shall skip some repetitive details.

At the phase boundary $R=R_0$, the field-strength components given by (\ref{R}) reduce to:
\begin{eqnarray}\label{R-}
	R^{01}&\doteq& 0,\nonumber\\
	R^{02}&\doteq&-\left(1-\frac{2M}{\rho}-\frac{\Lambda \rho^2}{3}\right)^{\frac{1}{2}}\Big(\frac{M}{\rho^2}-\frac{\Lambda \rho}{3}\Big)\dot{t} dT\wedge d\theta,\nonumber\\
		R^{03}&\doteq &
	-\left(1-\frac{2M}{\rho}-\frac{\Lambda \rho^2}{3}\right)^{\frac{1}{2}}\Big(\frac{M}{\rho^2}-\frac{\Lambda\rho}{3}\Big)\dot{t}\sin\theta dT\wedge d\phi \nonumber\\
	R^{12}	&\doteq& 
	-\left(1-\frac{2M}{\rho}-\frac{\Lambda \rho^2}{3}\right)^{-\frac{1}{2}}\Big(\frac{M}{\rho^2}-\frac{\Lambda \rho}{3}\Big)\dot{\rho} dT\wedge d\theta
	\nonumber\\
	R^{23}&\doteq & \Big(\frac{2M}{\rho}+\frac{\Lambda \rho^2}{3}\Big)\sin\theta d\theta\wedge d\phi\nonumber\\
	R^{31}&\doteq &
-\left(1-\frac{2M}{\rho}-\frac{\Lambda \rho^2}{3}\right)^{-\frac{1}{2}}\Big(\frac{M}{\rho^2}-\frac{\Lambda\rho}{3}\Big)\dot{\rho}\sin\theta dT\wedge d\phi\nonumber\\
&&~
\end{eqnarray}
Using the continuity constraints in (\ref{cont-R}) and (\ref{cont-g}) leads to:
\begin{eqnarray}
\dot{\rho}&\doteq & 0,\nonumber\\
F(T)&\doteq &\left(1-\frac{2M}{\rho}-\frac{\Lambda \rho^2}{3}\right)^{\frac{1}{2}}\dot{t},\nonumber\\
\lambda^2+\frac{\Lambda}{3}&\doteq& \frac{1}{\rho^2}\Big(1-\frac{2M}{\rho}\Big)
\end{eqnarray}
The non-static solution to the equation of motion now reads:
\begin{equation}\label{F-soln-}
	F(T) = \frac{\dot{\rho}}{\sqrt{\Big(\lambda^2+\frac{\Lambda}{3}\Big) \rho^2-1+\frac{2M}{\rho}}}
\end{equation}

Regarding the static bubble, the solutions for $\rho(T)$, $\lambda$ and $F$ read:
\begin{eqnarray}\label{lambda-}
\rho(T)&=&r_0=3M\nonumber\\
\lambda &=&\Big(\frac{1}{27 M^2}-\frac{\Lambda}{3}\Big)^{\frac{1}{2}},\nonumber\\
F(T)&=&\frac{1}{\sqrt{3}}\Big(1-9M^2\Big)^{\frac{1}{2}}\dot{t}(T)
\end{eqnarray}

\subsection*{Topological charge}

Note that even though the solutions for $\lambda$ and the metric obtained in (\ref{lambda-}) is different from the Schwarzschild case given by (\ref{lambda-sol}), the location of the phase boundary coincides with the photon sphere of the standard Schwarzschild-dS (AdS) black hole, exactly as earlier. The value of the topological charge (\ref{Q}) is found as:
\begin{eqnarray}
Q=(3\lambda^2+\Lambda)r_0^2=1
\end{eqnarray}
Thus, for both $\Lambda>0$ and $\Lambda<0$, the topological number equals the one obtained for the Schwarzshild bubble. 

Since the relation between $M$ and $Q$ remains the same as earlier, we rediscover the topological interpretation of the Schwarzschild `mass' $M$ here in presence of $\Lambda$.

\subsection*{Curvature scalars}
Following the analysis elucidated in the Schwarzschild case, the only non-vanishing components of the Riemann curvature tensor in the coordinate basis are displayed below:
\begin{eqnarray}
\hat{R}_{T\theta}^{~T\theta}&=&\lambda^2+\frac{\Lambda}{3}-\frac{M}{\rho^3}=\hat{R}_{T\phi}^{~T\phi},\nonumber\\
\hat{R}_{\theta\phi}^{~\theta\phi}&=&\lambda^2+\frac{\Lambda}{3}+\frac{2M}{\rho^3}
\end{eqnarray}
Since $\rho(T)\geq 3M$, these components are finite. These determine all possible curvature scalars, which are in turn finite.

Similarly, for the static solutions, the only nontrivial component is finite: $\hat{R}_{\theta\phi}^{~\theta\phi}=\frac{1}{9M^2}$.

\section{CONCLUDING REMARKS}
Our analysis begins with a general question, that is, whether the black hole `mass' could have an origin other than genuine matter or torsion. We approach this problem by exploring the possible ways to construct a bubble spacetime solution in vacuum by gluing together the $g=0$ and $g\neq 0$ phases of first-order gravity. These studies lead to a set of new spacetime solutions with a universal topological charge. This charge directly defines the bubble mass, which supercedes the mass of a conventional black hole. Further, the curvature scalars at the non-invertible phase are finite everywhere.

Remarkably, the equations of motion fix the location of the phase boundary uniquely for the static solutions. It necessarily coincides with the photon sphere of a standard black hole. The underlying topological charge obtained from a conserved current is an essential characteristic of this special surface. This may be contrasted with a phase boundary located at the event horizon, which corresponds to a vanishing charge. Thus, within our construction of the bubble spacetimes in vacuum, the photon sphere emerges as a more fundamental geometrical surface. Let us note that topological features of the photon sphere had also been studied earlier from a different perspective \cite{cunha,wei}. In contrast with our work here, these studies deal with standard Einsteinian black holes in the invertible metric phase and employ the construction of a nontrivial homotopy map. 

Our analysis and results motivate a number of future directions and potential applications. Whether these solutions could emerge as practical (regular) alternatives to the singular black holes remains an intriguing question.  It does seem worthwhile to pursue the idea of a topological alternative to the black hole mass in a broader context. One wonders whether the same feature could emerge also in a gravity theory with a gauge group higher than $SO(3,1)$ or in higher dimensions. Further, given that the degenerate phase of the bubble spacetime remains inaccessible to any outside observer following a timelike radial geodesic, these topological solutions to the field equations could provide a new perspective into the `dark matter' problem. These bubbles should be probed further for their possible signatures in the context of the ringdown phase of strong gravity phenomena and shadow observables, where the photon surface is known to play a critical role.

Some of the important questions above deserve a thorough investigation elsewhere.

\begin{acknowledgments}
The support of the ANRF, Govt. of India, through the ARG-MATRICS grant no. ANRF/ARGM/2025/000231/TS is gratefully acknowledged. It is a pleasure to thank Nemani V. Suryanarayana for rewarding discussions on this work.
\end{acknowledgments}

 \bibliography{topological-schw-mass.bib}

\end{document}